\documentclass[letterpaper,twocolumn,10pt]{article}
\usepackage{usenix2019_v3}
\usepackage{amsmath,amssymb}
\AddToHook{env/equation*/begin}{%
  \setlength{\abovedisplayshortskip}{\abovedisplayskip}%
  \setlength{\belowdisplayshortskip}{\belowdisplayskip}%
}
\usepackage{graphicx}
\usepackage{stfloats}
\makeatletter
\@ifundefined{Gread@transgrouptrue}{}{\let\Gread@transgrouptrue\Gread@transgroupfalse}
\makeatother
\usepackage{pifont}
\usepackage{booktabs,tabularx,array}
\usepackage{algorithm,algpseudocode}
\usepackage{multicol,textcomp}
\usepackage{xurl}
\hypersetup{hidelinks,pdftitle={RePrism: Joint Request and Representation Scheduling for LLM Agent Serving},pdfauthor={Xijiang Shen, Aoli Zhu, Yanbo Tan, Xiyuan Hu, Huailiang Tan}}

\date{}
\author{%
  Xijiang Shen\textsuperscript{1}\quad
  Aoli Zhu\textsuperscript{1}\quad
  Yanbo Tan\textsuperscript{2}\quad
  Xiyuan Hu\textsuperscript{1}\quad
  Huailiang Tan\textsuperscript{1,*}\\[3pt]
  {\normalsize\textsuperscript{1}Hunan University\qquad
    \textsuperscript{2}Shanghai Jiao Tong University}\\[2pt]
  {\small\href{mailto:xjshen@hnu.edu.cn}{xjshen@hnu.edu.cn}\quad
    \href{mailto:zhuaoli@hnu.edu.cn}{zhuaoli@hnu.edu.cn}\quad
    \href{mailto:bayernmunich520@sjtu.edu.cn}{bayernmunich520@sjtu.edu.cn}}\\
  {\small\href{mailto:huxiyuan556@gmail.com}{huxiyuan556@gmail.com}\quad
    \textsuperscript{*}\href{mailto:tanhuailiang@hnu.edu.cn}{tanhuailiang@hnu.edu.cn}
    (corresponding author)}%
}
\makeatletter
\renewcommand{\@maketitle}{%
  \newpage
  \begingroup
    \centering
    {\Large\bfseries\@title\par}
    \vspace{6pt}
    {\large\normalfont\begin{tabular}{@{}c@{}}\@author\end{tabular}\par}
  \endgroup
  \vspace{\baselineskip}%
}
\makeatother
\title{RePrism: Joint Request and Representation Scheduling for LLM Agent Serving}

\begin{document}
\maketitle
\thispagestyle{plain}

% word:11
\begin{abstract}
Agentic workflows often invoke large language models multiple times. When the dependencies are visible to the serving system, it can use them to schedule requests and cache reusable content, improving workflow execution efficiency. Prior work has largely focused on ordering requests or managing shared resources under a fixed caching scheme. A reusable prompt that specifies model behavior, however, can have multiple representations. We present RePrism, a serving system that jointly selects which ready request to serve and which eligible representation of its prompt to use. Its goal is to improve throughput and shorten session completion time (SCT). The scheduler ranks pairs of requests and representations by combining each pair’s immediate value with an estimate of its future value based on the visible workflow. The resource manager allocates GPU memory to both LoRA adapters and cached key-value states. It also coordinates their admission and eviction as memory pressure changes. We compare RePrism with S-LoRA on a dynamic workload with an arrival rate of 1.8 sessions per second, and it achieves 1.37\ensuremath{\times} the request throughput and reduces P95 SCT by a factor of 4.10. Under controlled load, RePrism reduces mean time to first token by a factor of 5.26 relative to S-LoRA. On a controlled workload, RePrism's joint scoring achieves 2.7\% higher request throughput and 12.3\% lower mean SCT than a policy that ranks requests before selecting their representations.
\end{abstract}

% word:12
\noindent \textbf{Keywords:} agent workflows; LLM serving; prompt representation; LoRA adapter; KV cache;

% word:13
\section{Introduction}\label{sec:1}

% word:14
The use of large language models (LLMs) now extends beyond single-turn question answering. To handle more complex tasks, a growing number of applications use agentic workflows to coordinate multiple LLM calls~\cite{ref01,ref02,ref03,ref04}. We focus on workflows in which the data and control dependencies among calls are known before execution or can be represented by the agent framework as a static or partially static directed acyclic graph (DAG) for each session~\cite{ref05,ref06,ref07}. A call can run only after its predecessors have completed and its data dependencies are satisfied. Users care about when the entire session finishes, so evaluating a serving system requires session-level metrics such as session completion time (SCT) alongside request latency and time to first token (TTFT)~\cite{ref08,ref09}. The visible workflow reveals dependencies and potential reuse, giving the scheduler information about future requests~\cite{ref10,ref11}.

% word:15
Context in agent workloads has different scopes of reuse. System or role prompts that specify behavioral rules and output formats are relatively stable, and they can be reused both within and across sessions~\cite{ref12,ref13}. Such fixed prompt can span hundreds or even thousands of tokens~\cite{ref10}, and context generated during execution may also be consumed by successor nodes. Reprocessing this content incurs repeated prefill costs~\cite{ref14,ref15}, while retaining all reusable cache consumes scarce GPU memory. The serving system must balance recomputation against memory cost.

% word:16
A prompt that specifies model behavior can have several representation methods. \textit{Direct} reprocesses the entire prompt for each request and retains no state for reuse. \textit{KV} retains the key-value states for the full prefix, avoiding recomputation when the same prefix recurs. \textit{Adapter} uses LoRA weights trained offline to elicit the behavior specified by the prompt. \textit{Hybrid} extends Adapter by also retaining the KV states for the residual context. The four representations trade off computation, storage, and the cost of acquiring state~\cite{ref16,ref17,ref18,ref19,ref20}. Direct and KV can be treated as exact execution paths. Adapter and Hybrid are parameterized approximations subject to behavioral quality requirements, with no guarantee of token-by-token agreement with the exact paths. Only a subset of the four representations may be eligible for a given request because some behaviors specified by prompts cannot be internalized in adapter parameters.

% word:17
Prior work has pursued three largely separate directions. Prefix caching systems reduce repeated prefill computation by optimizing what to cache, evict, and prefetch, and can use workflow structure to predict future reuse~\cite{ref11,ref21}. Multi-LoRA serving systems optimize batching, adapter residency, and GPU memory management~\cite{ref22,ref23}. Prompt internalization methods focus on constructing a parameterized representation of a fixed prompt offline so that the model exhibits the specified behavior during online inference~\cite{ref19}. These approaches primarily address execution and cache management for a given representation, or the offline construction of that representation. When several representations are available for the same request, the serving system must decide both which ready request to execute next and which representation best advances the workflow.

% word:18
Our key observation is that request scheduling and representation selection are tightly coupled. This coupling affects how the scheduler compares execution costs across requests. If it first ranks ready requests by the value of executing each one to its session and chooses a representation only after fixing the order, representation specified computation and resource costs cannot influence that ranking. The choice of representation also affects the resources available to later calls. Retaining a large KV cache may evict cache that a successor is about to reuse, whereas choosing the Adapter representation may require reprocessing residual inputs. A representation that lowers latency for the current request therefore not always shorten session completion time. To make an online decision, the scheduler must account for current execution costs, predicted reuse, and contention for shared GPU memory.

% word:20
\begin{figure}[!t]
\centering
\includegraphics[width=.73\linewidth]{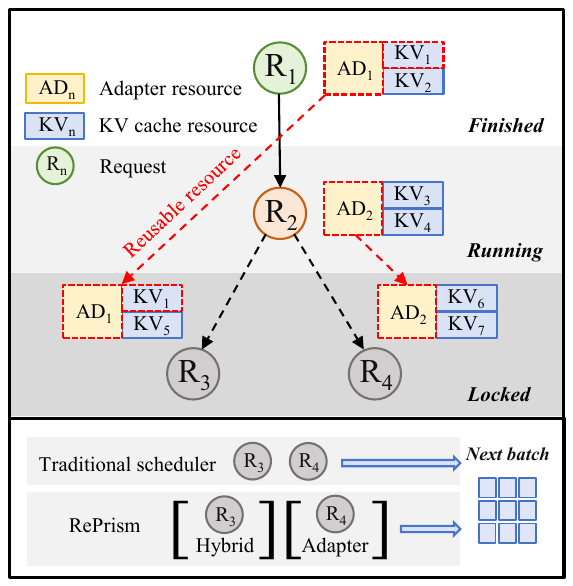}
\caption{Workflow reuse and batch selection. The DAG shows which cache resources later requests can reuse. RePrism pairs each request in the next batch with a prompt representation.}\label{fig:1}
\end{figure}

% word:21
We present RePrism, a system that jointly schedules requests and selects their representations for agentic workflows. Its goal is to sustain high request throughput and shorten SCT under a limited GPU memory budget. As illustrated in Figure~\ref{fig:1}, RePrism pairs each ready request with each of its eligible representations. A joint scheduler and a resource manager coordinate candidate selection and shared cache management to advance sessions. The scheduler uses the visible workflow structure to rank pairs, and the resource manager coordinates the admission and eviction of reusable resources.

% word:22
We implement RePrism on S-LoRA and evaluate it on an NVIDIA RTX 3090 GPU with Llama-2-7B-Chat. Our evaluation spans three application templates and includes both dynamic and controlled workloads. On the dynamic workload, at an arrival rate of 1.8 sessions per second, RePrism achieves 1.37\ensuremath{\times} the request throughput of S-LoRA and reduces P95 SCT by a factor of 4.10. On the controlled workload, RePrism reduces mean TTFT by a factor of 5.26 relative to S-LoRA. Compared with selecting representations after ranking requests, joint scoring raises request throughput by 2.7\% and lowers mean SCT by 12.3\%.

% word:23
This paper makes three contributions:

% word:24
\begin{itemize}
\item We characterize the trade-offs between computation and GPU memory across prompt representations and show how reuse patterns and memory pressure affect representation choice.

% word:25
\item We present RePrism, which uses workflow information to jointly select a ready request and an eligible representation of its prompt. RePrism manages reusable resources under a shared GPU memory budget.

% word:26
\item We evaluate RePrism against representative serving systems. Ablation and sensitivity studies characterize the benefits of joint scheduling and the conditions under which it is effective.
\end{itemize}

% word:27
\section{Background}\label{sec:2}

% word:28
\textbf{Agentic workflows and session objectives.} Traditional online LLM serving focuses on individual requests~\cite{ref16,ref24,ref25,ref26,ref27}. Agentic applications rely on several interdependent calls to accomplish one user goal~\cite{ref01,ref05,ref28}, and users always care about the completion of whole task instead of single request. We call the execution from task arrival to final output a \textit{session}. For session \(s\), the graph \(G_{s}=(V_{s},E_{s})\) captures the dependencies currently visible to the scheduler. Each node \(v\in V_{s}\) represents an LLM call or a logical execution stage. An edge \((u,v)\in E_{s}\) means that node \(v\) depends on node \(u\) for data or control. The visible workflow shows each node's position in the graph and helps the scheduler estimate future reuse~\cite{ref06,ref08,ref11}. Ready nodes differ in how they advance a session. Completing a node may unblock several successors or change the critical path through a conditional branch. When successor calls and their reusable context are known, the scheduler can anticipate reuse and decide which cached resources to retain or load.

% word:30
\begin{figure}[!t]
\centering
\includegraphics[width=.87\linewidth]{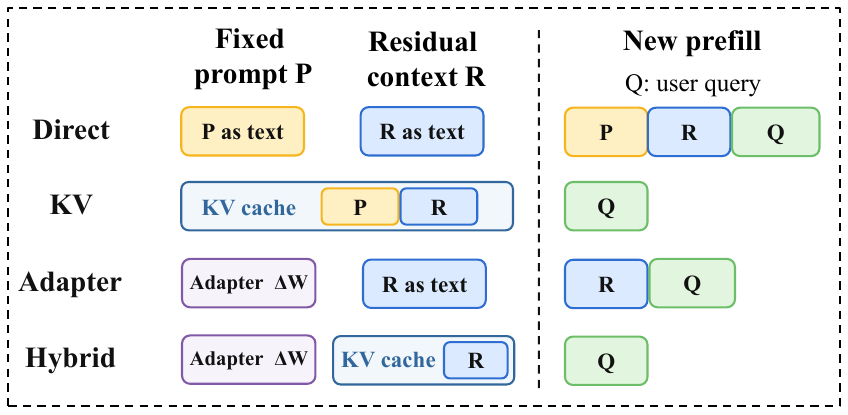}
\caption{Cached resources and prefill inputs for Direct, KV, Adapter, and Hybrid.}\label{fig:2}
\end{figure}

% word:31
\textbf{Prompt representation trade-offs.} A request in an agentic workflow typically includes a stable prefix and input that varies by session or call~\cite{ref10,ref12,ref13,ref17}. We decompose the input by scope of reuse as \(P\parallel R\parallel Q\). \(P\) is the system or role prompt that specifies model behavior and output format and can be reused across sessions. \(R\) is the context produced by the current session or execution path. Successors in the same session typically share it, and rarely reuse across sessions. \(Q\) is the call-specific query with little opportunity for reuse.

% word:32
Figure~\ref{fig:2} shows the four runtime representations. Direct processes \(P\parallel R\parallel Q\) and releases its cache after request completed. KV retains the key-value tensors computed during prefill of a shared prefix so that a matching request can resume at the suffix. Adapter uses offline prompt internalization to train LoRA weights \(\Delta W_{P}\) that elicit behavior similar to that induced by the explicit prompt~\cite{ref18}, and it only needs prefill \(R\parallel Q\) at runtime. Adapter size depends on tensor dimensions, including rank and the adapted layers, and does not scale directly with prompt length. Hybrid combines two reuse lifetimes. It represents \(P\) with an adapter and caches the residual KV tensors that the base model computes for \(R\) with that adapter loaded. On a hit for both, only \(Q\) remains to process. Compared with Adapter, Hybrid saves the cost of reprocessing \(R\) on a hit but incurs the cost of building and storing the residual KV cache. The benefit of a cache hit must therefore be weighed against loading or rebuilding costs and the memory left for other requests. Improving request throughput and shortening SCT under a limited GPU memory budget requires considering request order and its representation together.

% word:33
\section{Characterizing Prompt Representations}\label{sec:3}

% word:34
In this section, we describe behavioral qualification for Adapter and Hybrid, measure the computation and GPU memory costs of four representations, and evaluate proxy adapters for scheduling experiments. These results inform candidate construction and online representation selection in Section~\ref{sec:4}. Section~\ref{sec:5.1} describes the experimental setup.

% word:35
\subsection{Prompt Internalization and Behavioral Qualification}\label{sec:3.1}

\begin{figure*}[!t]
\centering
\includegraphics[width=.75\linewidth]{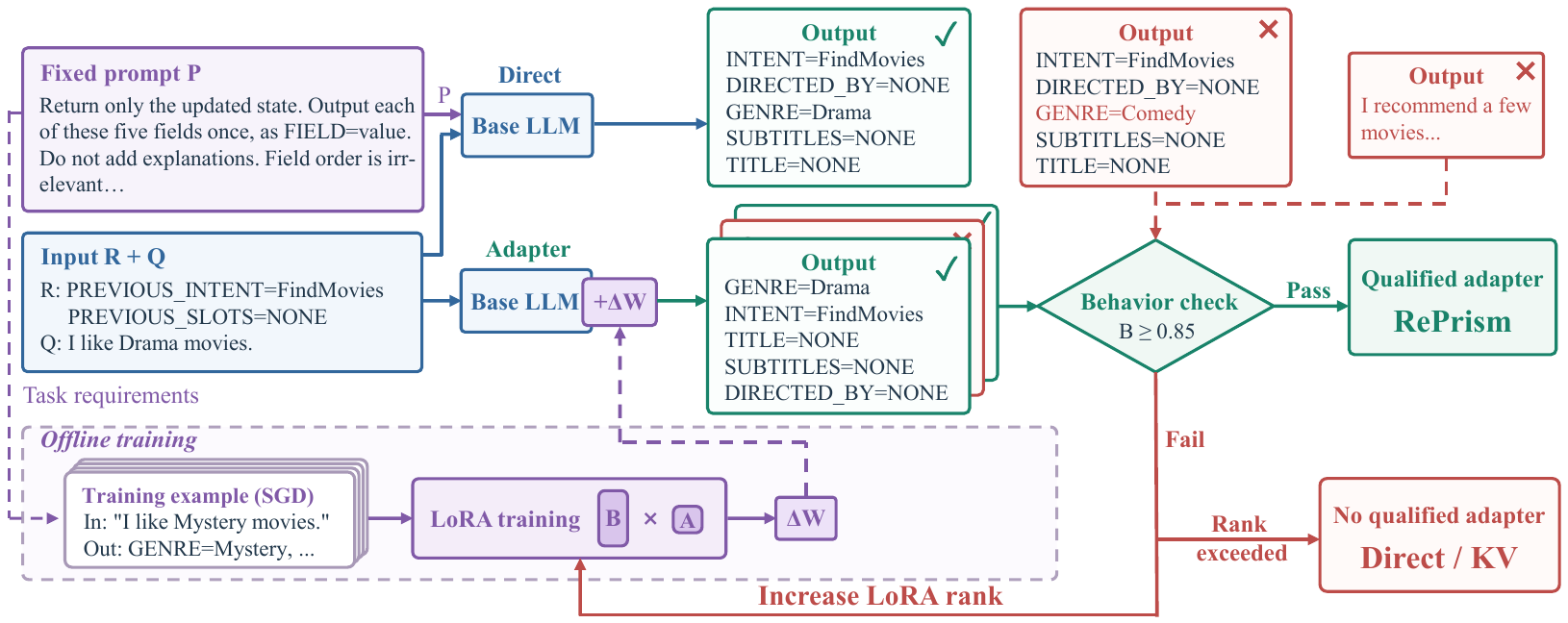}
\caption{Prompt internalization and behavioral qualification of LoRA adapters. An adapter qualifies for its prompt if its behavioral score is at least 0.85.}\label{fig:3}
\end{figure*}

% word:36
\textbf{Internalization objective.} Drawing on context distillation~\cite{ref20,ref29,ref30} and fixed-input parameterization~\cite{ref19,ref31}, we define two configurations of the base model. The \textit{teacher} takes \(P\) as an explicit input, and the \textit{student} omits \(P\) and uses trainable LoRA weights. As Figure~\ref{fig:3} shows, we train the adapter offline using the dataset's ground-truth answers and then check whether it meets the behavioral requirements for \(P\). On unseen inputs \(R\parallel Q\), the student should approximate the teacher's behavior on the corresponding \(P\parallel R\parallel Q\).

% word:39

% word:40
\textbf{Behavioral qualification.} We load the trained adapter into the base model and evaluate it on an independent test set. Its behavioral score is the fraction of outputs that meet the structural and behavioral requirements set before training. We construct fixed prompts from the Schema-Guided Dialogue (SGD) dataset~\cite{ref32}. For the Media\_1 prompt (1004 tokens), adapters of ranks 4 and 8 scored below the preset threshold of 0.85, while the adapter of rank 16 scored 0.917 and qualified. Higher ranks can encode more information, but length only partly reflects information content, so longer prompts need not require higher ranks. A base model's capabilities also constrain how it can respond to a prompt, so the same prompt may elicit entirely different behavior across models. For subsequent experiments, we pair adapter ranks 4, 8, 16, and 32 with fixed prompts \(P\) of 378, 733, 1004, and 1723 tokens, respectively. Section \ref{supp:training} of the supplementary material gives training details and results.

% word:42
\subsection{Latency and GPU Memory Use}\label{sec:3.2}

% word:43
For the prompt \(P\) with qualified adapters, we compare the computation and caching costs of serving the same request with Direct, KV, Adapter, or Hybrid.

\begin{figure*}[!t]
\centering
\includegraphics[width=.87\linewidth]{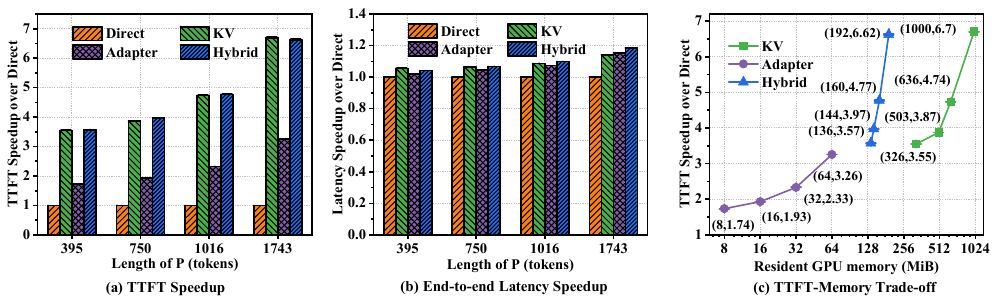}
\caption{Warm-path performance and GPU memory use across prompt representations. Panels (a) and (b) show TTFT speedup and end-to-end latency speedup, respectively, as prompt length varies. Panel (c) shows TTFT speedup versus resident GPU memory. All speedups are relative to Direct.}\label{fig:4}
\end{figure*}

% word:44
Figure~\ref{fig:4} compares the four representations on the warm path as the length of \(P\) varies, and the reported length of \(P\) includes the model's chat template. \(R\), \(Q\), and \(O\) are 256, 256, and 128 tokens long, respectively. Direct is the baseline in all four configurations. KV, Adapter, and Hybrid each avoid some repeated prefill of \(P\parallel R\), with a peak TTFT speedup of over 6\ensuremath{\times} in Figure~\ref{fig:4}(a). Increasing the length of \(P\) raises Direct's prefill cost and the benefit of representation selection. Figure~\ref{fig:4}(b) shows the output generation limits the end-to-end latency speedup to about 1.18\ensuremath{\times} at best across the tested lengths of \(P\). Adapter and Hybrid have higher end-to-end latency than KV for short \(P\) but become faster as \(P\) grows. With KV, each decoding step attends to all preceding tokens. Adapter and Hybrid encode \(P\) in LoRA weights, leaving fewer tokens to attend to.

% word:45
These computational gains come at different GPU memory costs. For the longest prompt in Figure~\ref{fig:4}(c), KV uses 1000 MiB of GPU memory, compared with 64 MiB for Adapter and 192 MiB for Hybrid. Adapter uses the least memory, but its TTFT speedup is only about half that of either KV or Hybrid. KV achieves the greatest TTFT speedup, but it uses about five times as much memory as Hybrid and has higher end-to-end latency than both Adapter and Hybrid. Hybrid thus uses less GPU memory than KV and achieves a greater TTFT speedup than Adapter. We measured the GPU memory each representation occupies when registered on its own, but these values exclude savings from sharing across requests. For example, when requests share \(P\) but have different \(R\), both KV and Hybrid can reuse what they retain for \(P\) and rebuild only the part associated with \(R\).

% word:46
Hybrid does not always outperform Adapter. Its warm-path savings must be weighed against the cost of building and retaining the residual KV cache. Consider a rank-16 adapter, a 256-token query (\(Q\)), and a 128-token output (\(O\)). For residual contexts of 64, 128, 256, and 512 tokens, it takes 25.28, 37.33, 65.47, and 140.79 ms, respectively, to build and register the residual KV cache for the first time. The corresponding cache resources occupy an additional 32, 64, 128, and 256 MiB of GPU memory while resident. At those same lengths, each subsequent hit saves 29.77, 33.69, 70.18, and 120.25 ms of processing time relative to Adapter. Repeated hits can quickly amortize the registration cost, but the resources continue to occupy memory between hits. Under memory pressure, evictions and contention among concurrent requests can reduce throughput or increase tail latency. Warm-path latency alone is therefore an insufficient basis for choosing Hybrid. The choice depends on how much processing time subsequent hits save while the resources remain resident.

% word:48

% word:49
\subsection{Validating Proxy Adapters for Scheduling Experiments}\label{sec:3.3}

% word:50
To study these trade-offs across more adapters and ranks in the scheduling experiments, we use proxy adapters with the same tensor structures as the semantic adapters trained in Section~\ref{sec:3.1} but different LoRA weights. Prior work on multi-LoRA serving also constructs performance workloads from tensor shapes~\cite{ref33}. To validate this substitution, we compare a qualified rank-16 adapter with its proxy using identical inputs and runtime settings. For both Adapter and Hybrid, the relative differences in mean throughput and mean P95 SCT between the two adapters are below 1.3\%. Switching from Adapter to Hybrid increases throughput by 8.72\% with the semantic adapter and 10.26\% with the proxy, and reduces P95 SCT by 8.03\% and 9.32\%, respectively. The two sets of weights yield similar execution performance and similar gains from Hybrid. This agreement supports using proxy adapters to study RePrism's scheduling behavior in subsequent experiments.

% word:51
\section{Joint Request Scheduling and Representation Selection}\label{sec:4}

% word:52
\subsection{RePrism Architecture}\label{sec:4.1}

% word:54
\begin{figure}[!t]
\centering
\includegraphics[width=.88\linewidth]{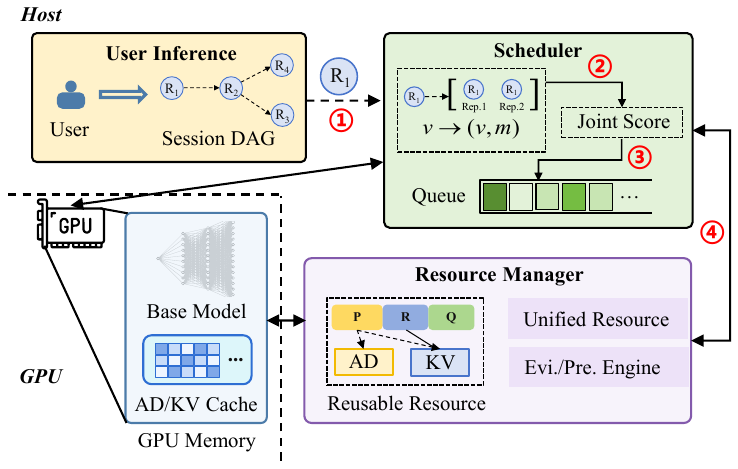}
\caption{RePrism architecture. The scheduler jointly selects requests and their representations, and the Resource Manager coordinates cache admission and eviction.}\label{fig:5}
\end{figure}

% word:55
RePrism jointly schedules requests and selects their representations under a shared GPU memory budget. Figure~\ref{fig:5} shows its two components, the \textit{Joint Scheduler} and the \textit{Resource Manager}. We assume that when a session's first request arrives, the scheduler has access to the visible workflow state and request attributes. The scheduler first identifies requests whose dependencies are satisfied\ding{172}. Then it pairs each with representations that meet the behavioral requirements to form candidates for joint scoring\ding{173}, and prioritizes them by their contribution to session progress and their representation costs\ding{174}. The resource manager reports which resources are resident and estimates GPU memory requirements. Once a batch is selected for admission, it coordinates resource loading and necessary evictions\ding{175}. We use a single GPU in this work, but RePrism could be extended to serve larger models with tensor or pipeline parallelism across GPUs.

% word:56
\subsection{Pairing Ready Requests with Eligible Representations}\label{sec:4.2}

% word:57
Conventional multi-LoRA schedulers~\cite{ref22,ref34,ref35} take an individual request \(v\) as the unit of scheduling, typically without tracking workflow structures. In an agentic workflow, a request node exposed to scheduler can run only when its required predecessors have finished and its branch conditions are met~\cite{ref08,ref36,ref37}. A submitted request therefore enters the ready set \(\mathcal{R}_{t}\) at scheduling time \(t\) only if these conditions hold. For each ready request \(v\), the scheduler checks which representations in \(\mathcal{M}(v)\subseteq \{\mathrm{Direct},\mathrm{KV},\mathrm{Adapter},\mathrm{Hybrid}\}\) are eligible. Adapter and Hybrid are eligible only if the fix prompt can be represented by an adapter that passes the validation. Let \(E_{t}(v,m)\) be the product of two indicators, one for satisfied request dependencies and one for representation eligibility. The set of candidates for online scheduling is:
% word:58
\[
\mathcal{C}_{t}=\{(v,m)\mid v\in \mathcal{R}_{t},\;m\in \mathcal{M}(v),\;E_{t}(v,m)=1\}.
\]

% word:59
RePrism checks representation eligibility separately from cache residency. For example, Hybrid remains eligible if its residual KV cache is absent but can be built with the corresponding adapter. However, even when the residual context is the same, a model cannot directly reuse a residual KV cache built with a different adapter. Offline qualification thus determines which representations a request can use. Cache residency determines the cost of using them, and joint scoring determines which candidates the scheduler selects for execution.

% word:60
\subsection{Joint Scoring and Batch Selection}\label{sec:4.3}

% word:61
The RePrism scheduler ranks candidates for execution, and its design draws on multi-LoRA schedulers~\cite{ref22,ref34,ref35,ref38} and serving systems for agentic workflows~\cite{ref10,ref21,ref39}. Scoring the value of executing request node \(v\) and the value of its representation separately can yield a very different execution order from scoring them jointly. For example, a node on a longer remaining path may need extra time to build or load resources for representation \(m\), while a node that contributes less to session progress can reuse a resident cache entry immediately. A scheduler that fixes the request order before choosing how to represent each reusable context has little opportunity to weigh these two effects across requests.

\begin{figure*}[!t]
\centering
\includegraphics[width=.98\linewidth]{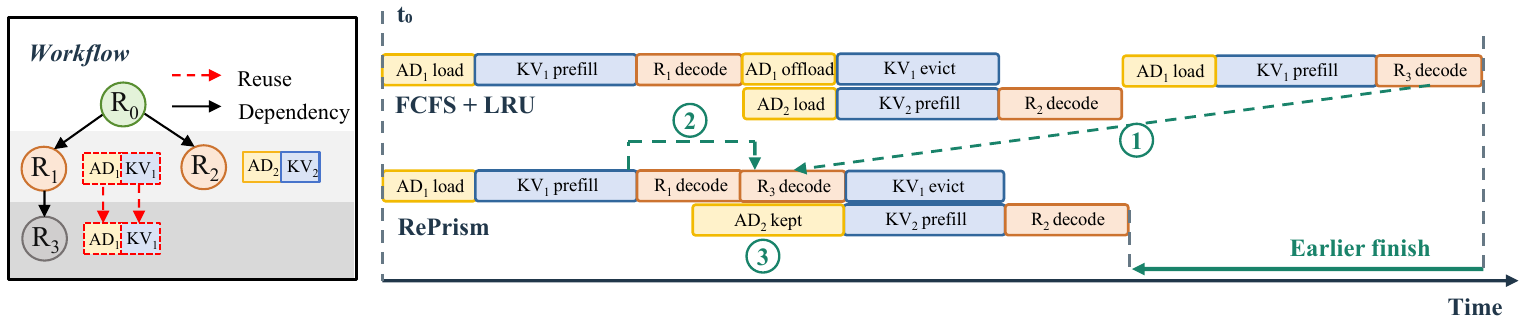}
\caption{Schematic execution timelines for FCFS with LRU eviction and for RePrism on the workflow at left. Scheduling \(R_3\) ahead of \(R_2\) lets \(R_3\) reuse the adapter and KV cache from \(R_1\).}\label{fig:6}
\end{figure*}

% word:62
Figure~\ref{fig:6} illustrates how RePrism completes a simple workflow sooner than FCFS scheduling with LRU eviction. The timelines for requests \(R_1\), \(R_2\), and \(R_3\) omit the prefill of \(Q\). Joint scoring moves \(R_3\) ahead of \(R_2\) because running \(R_2\) first would displace resources already in GPU memory\ding{172}. RePrism uses the visible workflow to choose representations that let \(R_3\) reuse the adapter and KV cache used by \(R_1\)\ding{173}. Since \(R_2\) will run soon, RePrism gives priority to retaining the resident resources it will need, avoiding a later reload\ding{174}. This section explains how RePrism implements \ding{172} and \ding{173}. Section~\ref{sec:4.4} covers \ding{174}.

% word:64

% word:65
\textbf{Joint score.} For each candidate \((v,m)\in \mathcal{C}_{t}\), RePrism computes a joint score by normalizing the node execution value \(V_{t}^{\mathrm{node}}(s,v)\) and the representation value \(V_{t}^{\mathrm{repr}}(v,m)\) separately, then multiplying them. We set the lower bound of the normalized range to 0.5 to prevent either term from overwhelming the other when its value is small:
% word:66
\[
S_{t}(s,v,m)=N_{[0.5,1]}(V_{t}^{\mathrm{node}}(s,v))\cdot N_{[0.5,1]}(V_{t}^{\mathrm{repr}}(v,m)).
\]

% word:67
Node priority combines the benefit of advancing the session with the time it has spent waiting to be scheduled:
% word:68
\[
V_{t}^{\mathrm{node}}(s,v)=\alpha _{p}P_{t}(s,v)+\alpha _{w}W_{t}(s),
\]
% word:69
where \(P_{t}(s,v)\) represents the value of advancing the current node \(v\) to help the session \(s\) finish sooner. RePrism uses the remaining token count to estimate the time needed to complete session \(s\). This estimate and the path criticality of node \(v\) together determine \(P_{t}(s,v)\). Path criticality refers to the node's contribution to unlocking downstream branches, unblocking a fork, or completing the session directly. \(W_{t}(s)\) represents waiting urgency and increases with the total time session \(s\) has spent waiting to be scheduled. \(W_{t}(s)\) excludes the candidate's execution time and the time spent waiting for parallel branches. This prevents a session from continually gaining priority simply because it takes longer in execution.

% word:70
To compare the benefits of executing a request with different representations, RePrism expresses the value of each representation in terms of its immediate execution benefit, its future reuse benefit, and the opportunity cost of the resources it uses:
% word:71
\[
\begin{aligned}V_{t}^{\mathrm{repr}}(v,m)= {}&\beta _{v}V_{t}^{\mathrm{now}}(v,m)+\beta _{g}G_{t}^{\mathrm{reuse}}(v,m)\\ &{}-\beta _{o}O_{t}^{\mathrm{resource}}(v,m).\end{aligned}
\]
% word:72
where \(V_{t}^{\mathrm{now}}(v,m)\) represents the immediate time savings from using \(m\), \(G_{t}^{\mathrm{reuse}}(v,m)\) represents the expected savings from future reuse, and \(O_{t}^{\mathrm{resource}}(v,m)\) represents the change in expected rebuild cost due to eviction. All three terms are expressed relative to the estimated remaining execution time of request \(v\) under Direct. We set all scoring weights to 1. Sections \ref{supp:progress} and \ref{supp:scoring} of the supplementary material give the full scoring rules.

% word:73
Predicting costs and benefits. RePrism builds a prediction model from offline measurements and applies it at runtime using each request’s parameters. It compares the estimated execution times under the representation \(m\) and Direct, accounting for any additional time needed to load or build resources for \(m\). To predict reuse, RePrism traverses the visible workflow to identify later requests that could reuse \(m\). It weights each request’s contribution by reuse distance and branch probability, giving more weight to reuse that is likely to occur soon. RePrism also estimates which resident cache resources the candidate would displace and the additional time needed to rebuild them, discouraging choices that require costly evictions.

% word:74
\textbf{Scheduling procedure.} Algorithm~\ref{alg:schedule} outlines the scheduling procedure given the visible workflow state \(G_t\), the running batch \(B_t\), and the GPU memory state \(D_t\).

% word:76
% 从当前 Word 的矢量算法图逐行转写, 隐藏源图中未显示的结束关键字.
\begin{algorithm}[!htbp]
\makeatletter
\renewcommand{\fnum@algorithm}{Algorithm~\thealgorithm:}
\makeatother
\caption{Joint Request Scheduling and Representation Selection}\label{alg:schedule}
\begingroup
\algtext*{EndFor}
\algtext*{EndIf}
\algrenewcommand{\algorithmicforall}{\textbf{for each}}
\algrenewcommand{\alglinenumber}[1]{\footnotesize #1}
\begin{algorithmic}[1]
\Statex \textbf{Input:} $G_t$, $B_t$, $D_t$
\Statex \textbf{Output:} $\Pi_t$
\State $\mathrm{UpdateState}(G_t,D_t)$
\State $\mathcal{R}_t \gets \mathrm{ReadyRequests}(G_t)$
\State $\mathcal{C}_t \gets \{(v,m) \mid v\in\mathcal{R}_t,\ m\in\mathcal{M}(v),\ E_t(v,m)=1\}$
\State $\mathcal{C}_t \gets \mathrm{FilterFeasible}(\mathcal{C}_t,B_t,D_t)$
\ForAll{$c\in \mathcal{C}_t$}
  \State $\mathrm{score}(c) \gets \mathrm{JointScore}(c)$
\EndFor
\State $\mathcal{C}_t \gets \mathrm{SortCandidates}(\mathcal{C}_t)$
\State $\Pi_t \gets \mathrm{EmptyPlan}()$
\ForAll{$c\in \mathcal{C}_t$}
  \State \textbf{if} $\mathrm{Request}(c)\in\Pi_t$ \textbf{then continue}
  \State $\mathrm{ok} \gets \mathrm{ProjectAdmit}(c,\Pi_t,B_t,D_t)$
  \State \textbf{if} $\mathrm{ok}$ \textbf{then} $\mathrm{Add}(\Pi_t,c)$
  \State \textbf{if} $\mathrm{BatchFull}(\Pi_t)$ \textbf{then break}
\EndFor
\If{$\Pi_t\ne\emptyset$}
  \State $\mathrm{ok},V \gets \mathrm{CommitAdmission}(\Pi_t,B_t,D_t)$ \enspace // Alg.~\ref{alg:admission}
  \State \textbf{if not} $\mathrm{ok}$ \textbf{then return} $\emptyset$
  \State $\Pi_t.\mathrm{victims} \gets V$
\ElsIf{$\mathrm{IdleSafePoint}()$}
  \State $\Pi_t.\mathrm{prefetch} \gets \mathrm{SelectPrefetch}(G_t,D_t)$
\EndIf
\State \Return{} $\Pi_t$
\end{algorithmic}
\endgroup
\end{algorithm}

The scheduler first filters out candidates that exceed the GPU memory budget or use an ineligible representation. It ranks the remaining candidates by their joint scores, which it recomputes at the start of each round because some terms depend on runtime conditions. For example, \(V_{t}^{\mathrm{now}}(v,m)\) depends on whether the cache needed for \(m\) is resident. To limit waiting, requests that have waited more than 60 s take precedence over other requests ranked by score, but subject to dependency and resource constraints. The ranking stays fixed within a round. Once the scheduler adds a request’s first feasible candidate in the ranking to the tentative batch, it skips all remaining candidates for that request.

% word:75
Each time the scheduler adds a candidate to the tentative batch, it updates the remaining GPU memory budget. Once the batch is assembled, the scheduler calls \texttt{CommitAdmission()} to decide whether to admit it and which resources to evict when necessary. During batch formation, a long, high-scoring request that does not fit in the remaining memory could block shorter requests behind it and leave GPU memory unused. To avoid this, the scheduler lets later requests bypass higher-scoring candidates that do not fit.

% word:77
\subsection{GPU Memory Management across Representations}\label{sec:4.4}

% word:78
To support the batch plan produced by joint scoring in Algorithm~\ref{alg:schedule}, Resource Manager allocates GPU memory across representations and handles resource admission and eviction.

% word:79
\textbf{Unified resource accounting.} Different representations require different GPU memory, and the joint score weighs the benefit of reuse against this memory cost. To support reuse, the Resource Manager maintains a variable-size region in GPU memory. It also reserves space for the dynamic KV cache that grows as the merged batch runs. Using the input and expected output lengths of all requests in the merged batch, the manager estimates \(T_{\mathrm{peak}}\), the peak number of tokens that require dynamic KV storage during prefill and decoding. Cached resources count only toward the cache region, and an entry shared by multiple requests counts only once. Every batch merge must satisfy:
% word:80
\[
M_{\mathrm{keep}}+M_{\mathrm{new}}+b_{\mathrm{KV}}T_{\mathrm{peak}}\le B_{\mathrm{runtime}}.
\]
% word:81
where \(M_{\mathrm{keep}}\) and \(M_{\mathrm{new}}\) are the cache bytes required by the running and proposed batches, respectively.\newline \(b_{\mathrm{KV}}\) is the dynamic KV cache per token in bytes, and \(B_{\mathrm{runtime}}\) is the runtime GPU memory budget. The Resource Manager allocates the remaining space to the cache region to retain resources for reuse by later requests. Section \ref{supp:peak} of the supplementary material describes how RePrism calculates the peak dynamic KV requirement.

% word:82
\textbf{Batch resource admission.} RePrism checks individual candidates before committing the batch. During candidate checks, the batch proposed by the scheduler remains tentative. \texttt{ProjectAdmit()} in Algorithm~\ref{alg:schedule} checks whether the new batch can be merged within the GPU memory budget, without evicting any resident cache resources. \texttt{CommitAdmission()} in Algorithm~\ref{alg:admission} makes the final decision. It will rejects the batch if its memory requirements exceed the budget or if evicting all eligible resources would still leave too little space for the batch. Otherwise, it selects enough resources for eviction and commits the resource reservations and evictions together.

% word:83
\begin{algorithm}[!htbp]
\makeatletter
\renewcommand{\fnum@algorithm}{Algorithm~\thealgorithm:}
\makeatother
\caption{Batch Resource Admission and Eviction}\label{alg:admission}
\begingroup
\algtext*{EndIf}
\algrenewcommand{\alglinenumber}[1]{\footnotesize #1}
\begin{algorithmic}[1]
\Statex \textbf{Input:} $\Pi_t$, $B_t$, $D_t$
\Statex \textbf{Output:} (success, victims)
\State $\mathcal{U} \gets \mathrm{UniqueResources}(\Pi_t)$
\State $T_{\mathrm{peak}} \gets \mathrm{ProjectPeak}(B_t,\Pi_t)$
\If{$\mathrm{HardInfeasible}(\mathcal{U},T_{\mathrm{peak}},D_t)$}
  \State \Return{} $\mathrm{false},\emptyset$
\EndIf
\State $b \gets \mathrm{NeedBytes}(\mathcal{U},T_{\mathrm{peak}},D_t)$
\State $V \gets \emptyset$
\If{$b>0$}
  \State $H \gets \mathrm{Protected}(B_t,\Pi_t,D_t)$
  \State $V \gets \mathrm{SelectVictims}(\mathrm{Evictable}(D_t)\setminus H,b)$
  \If{$\mathrm{Freed}(V)<b$}
    \State \Return{} $\mathrm{false},\emptyset$
  \EndIf
\EndIf
\State $\mathrm{AtomicReserveWithEvict}(\Pi_t,\mathcal{U},T_{\mathrm{peak}},V,D_t)$
\State \Return{} $\mathrm{true},V$
\end{algorithmic}
\endgroup
\end{algorithm}

% word:84
Cache eviction. The manager evicts resident cache only when directly adding a new candidate to the running batch would exceed the GPU memory budget. It then selects resources for eviction in score order. The manager marks a resource as Protected if a running request holds it, it is being loaded, or the current candidate needs it. \texttt{SelectVictims()} in Algorithm~\ref{alg:admission} excludes protected resources and selects which resources to evict in the current round. Eviction in workflows requires more than tracking recent accesses. A cost-aware policy must account for both the uneven popularity of cache resources and the cost of restoring them~\cite{ref10}. RePrism ranks resources for eviction by their retention value per byte, \(D_{i}\), defined as follows:
% word:85
\[
D_{i}=\frac{C_{i}}{B_{i}}\cdot (1+r_{i})\cdot (1+\frac{\min(4,h_{i})}{4})
\]
% word:86
where \(C_{i}\) is the rebuild time saved by retaining resource \(i\), as measured by profiling, and \(B_{i}\) is the number of bytes it occupies in GPU memory. Their ratio gives a base priority derived from measured costs. RePrism multiplies this priority by two factors based on the visible workflow and recent rebuild history. The first uses \(r_{i}\), a value in [0, 1] that estimates future demand for resource \(i\) from the order in which Algorithm~\ref{alg:schedule} ranks candidates by their joint scores. The second uses \(h_{i}\), the number of times resource \(i\) was rebuilt within 10 s of eviction. RePrism keeps each rebuild record for 60 s to reduce cache thrashing.

% word:87
The manager ranks cache resources by their estimated retention value per byte of GPU memory and selects lower-value resources for eviction first. Following this order alone can free much more memory than needed. Once the manager has selected enough resources, it will revisit them in decreasing order of retention value. It removes a resource from the eviction set whenever evicting the rest would still free enough memory. The manager offloads evicted adapters to CPU memory and reloads them from there on their next use, and it discards evicted KV caches and rebuilds them on their next use. Under our test conditions, rebuilding the KV cache took less time than transferring it from CPU to GPU memory. When no batch is running and the waiting queue is empty, RePrism can also prefetch representations that future requests may use. But this path is not triggered in our experiments.

% word:88
\section{Evaluation}\label{sec:5}

% word:89
\subsection{Experimental Setup}\label{sec:5.1}

% word:90
\textbf{Platforms.} We use one NVIDIA RTX 3090 GPU with 24 GB of memory, an Intel 8171M CPU configured with 52 logical cores, and 128 GB of host memory. We implement RePrism on S-LoRA~\cite{ref22} and use Llama-2-7B-Chat as the base model in the main experiments.

% word:91
\textbf{Resource budgets.} We set the peak memory budget for the entire GPU to 22000 MiB for all systems, allowing 3\% runtime variation. To account for differences in static GPU memory use across frameworks, we set the dynamic runtime budget to 14000 KV token-equivalents for Helium and 16000 for the other systems. Each token requires 0.5 MiB on Llama-2. These budgets cover reusable state and the dynamic KV cache, and total GPU memory use also includes memory for model weights and workspaces. We keep the model configuration fixed within each group of experiments. Section~\ref{sec:5.5.1} gives the configurations used in the cross-model experiments.

% word:92
\textbf{Baselines.} We compare RePrism with S-LoRA, Chameleon~\cite{ref34}, SGLang~\cite{ref17}, and Helium~\cite{ref21}. Table~\ref{tab:baselines} lists the representations and workflow information available to each system. S-LoRA and Chameleon can use matching adapters, but neither can use workflow information. Chameleon can also cache adapters in spare GPU memory. SGLang can retain previously computed KV states for reuse without workflow information. Helium uses workflow information to optimize KV reuse. All systems process identical workflows and differ only in their available representations and scheduling logic. Section~\ref{sec:5.4} further compares scheduling policies using the same execution backend and resource management policy.

% word:93
% 单栏三线表. 中间两列留给完整的表示名称和工作流可见性标签.
\begin{table}[tbp]
\setlength{\belowcaptionskip}{4pt}
\caption{Serving mechanisms in our evaluation. DAG access means the scheduler can use workflow dependencies.}\label{tab:baselines}
\centering
\setlength{\tabcolsep}{2.9pt}
\begin{tabularx}{\columnwidth}{@{}l>{\raggedright\arraybackslash}p{.278\columnwidth}>{\centering\arraybackslash}p{.112\columnwidth}>{\raggedright\arraybackslash}X@{}}
\toprule
\textbf{System} & \textbf{Prompt representations} & \textbf{DAG access} & \textbf{Serving mechanism} \\
\midrule
S-LoRA & Direct, Adapter & No & Batches requests across adapters \\
\addlinespace[2.5pt]
Chameleon & Direct, Adapter & No & Caches adapters and considers adapter size when scheduling \\
\addlinespace[2.5pt]
SGLang & Direct, KV & No & Reuses KV states for shared prefixes \\
\addlinespace[2.5pt]
Helium & Direct, KV & Yes & Uses the workflow to plan KV reuse \\
\addlinespace[2.5pt]
RePrism & Direct, KV, Adapter, Hybrid & Yes & Jointly selects requests and representations \\
\bottomrule
\end{tabularx}
\end{table}

% word:95
\textbf{Workloads.} A full evaluation of RePrism’s joint scheduling requires a dataset that captures workflow dependencies, prefix sharing across requests, and the mapping from prompts to adapters. Public production traces typically report only request arrival times and input and output token counts. For example, Preble~\cite{ref13} notes that the Azure trace~\cite{ref40} omits the prompt text, making it impossible to reconstruct which requests share a prefix.

\begin{figure}[!t]
\centering
\includegraphics[width=\linewidth]{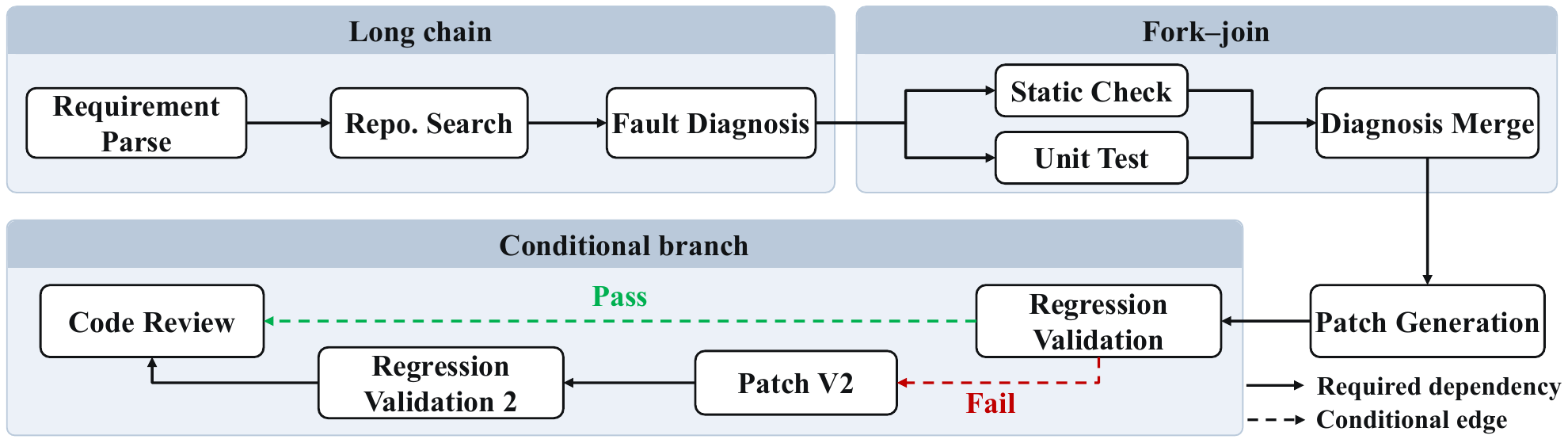}\par
\caption{CI repair workflow for a coding agent. Static checking and unit testing run in parallel before their results are merged. A failed regression check triggers a second patch and validation step.}\label{fig:7}
\end{figure}

% word:96
We design workloads with dynamic arrivals to evaluate end-to-end performance and controlled workloads to measure the gains from individual components. The controlled workloads in Sections~\ref{sec:5.3} through~\ref{sec:5.5} use the CI repair workflow shown in Figure~\ref{fig:7}. The dynamic workload in Section~\ref{sec:5.2} mixes CI repair, incident response, and multi-document analysis sessions. The latter two workflow templates contain eight and nine request nodes, respectively. All systems use the same workload configuration for a given experiment. The client sets the arrival time of each session's root request and releases successor requests as their dependencies are satisfied. It keeps at most 128 requests in flight and queues additional ready requests locally.

% word:98

% word:105
\begin{figure*}[!t]
\centering
\includegraphics[width=.90\linewidth]{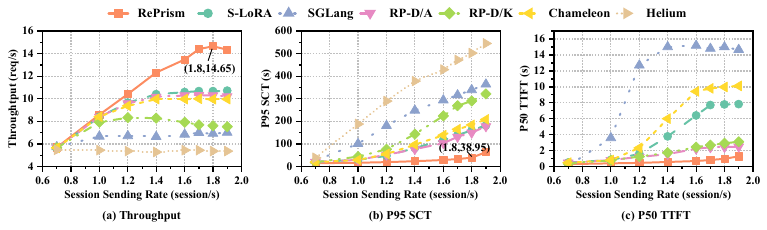}
\caption{End-to-end performance as the session arrival rate increases. Panels show (a) request throughput, (b) P95 SCT, and (c) P50 TTFT. Per-request TTFT is unavailable for Helium.}\label{fig:8}
\end{figure*}

% word:100
In the dynamic workloads, sessions arrive according to a Poisson process~\cite{ref27,ref34} at rates ranging from 0.7 to 1.9 sessions/s. The client submits new sessions for 200 s, then stops submitting sessions and lets the system run until all submitted requests are completed. Prompt \(P\) ranges from 409 to 1949 tokens, with lengths concentrated in four bands corresponding to the prompt lengths paired with the four adapter ranks in Section~\ref{sec:3.1}. Within each rank, we assign adapters to requests according to a Zipf distribution with exponent 1.1~\cite{ref33}. Residual context \(R\) ranges from 149 to 207 tokens. Query length \(Q\) and expected output length \(O\) vary by task template, with most values between 10 and 100 tokens.

% word:101
The controlled experiments comprise resource pressure tests and policy comparisons. The resource pressure tests in Section~\ref{sec:5.3} release all sessions at once to create high concurrency. For policy comparisons in Sections~\ref{sec:5.4} through~\ref{sec:5.5}, the client uses the same session arrival sequence for all policies. The interarrival times follow a gamma distribution with a shape parameter of 0.5, and the target arrival rate is 0.2 sessions/s. After all sessions have been submitted, each run continues until they finish. Each controlled workload contains 96 sessions and 1,056 requests. By default, the lengths of \(P\), \(R\), \(Q\), and \(O\) are 1152, 256, 256, and 128 tokens, respectively, and the workload provides a pool of 32 rank-16 adapters for use in place of \(P\).

% word:102
\textbf{Metrics.} We report request throughput, session completion time (SCT), and time to first token (TTFT). Throughput is the number of completed requests divided by the total time from when the client starts sending requests until the server drains all requests. SCT is the time from a session’s scheduled arrival until its target node completes. For controlled workloads, we primarily report mean SCT to measure overall completion efficiency. For dynamic workloads, we primarily report P95 SCT to measure session tail latency. All figures report results averaged over five runs. TTFT is the time from when the client submits a request until it receives the first response containing a token.

% word:112
\begin{figure}[!t]
\centering
\includegraphics[width=1\linewidth]{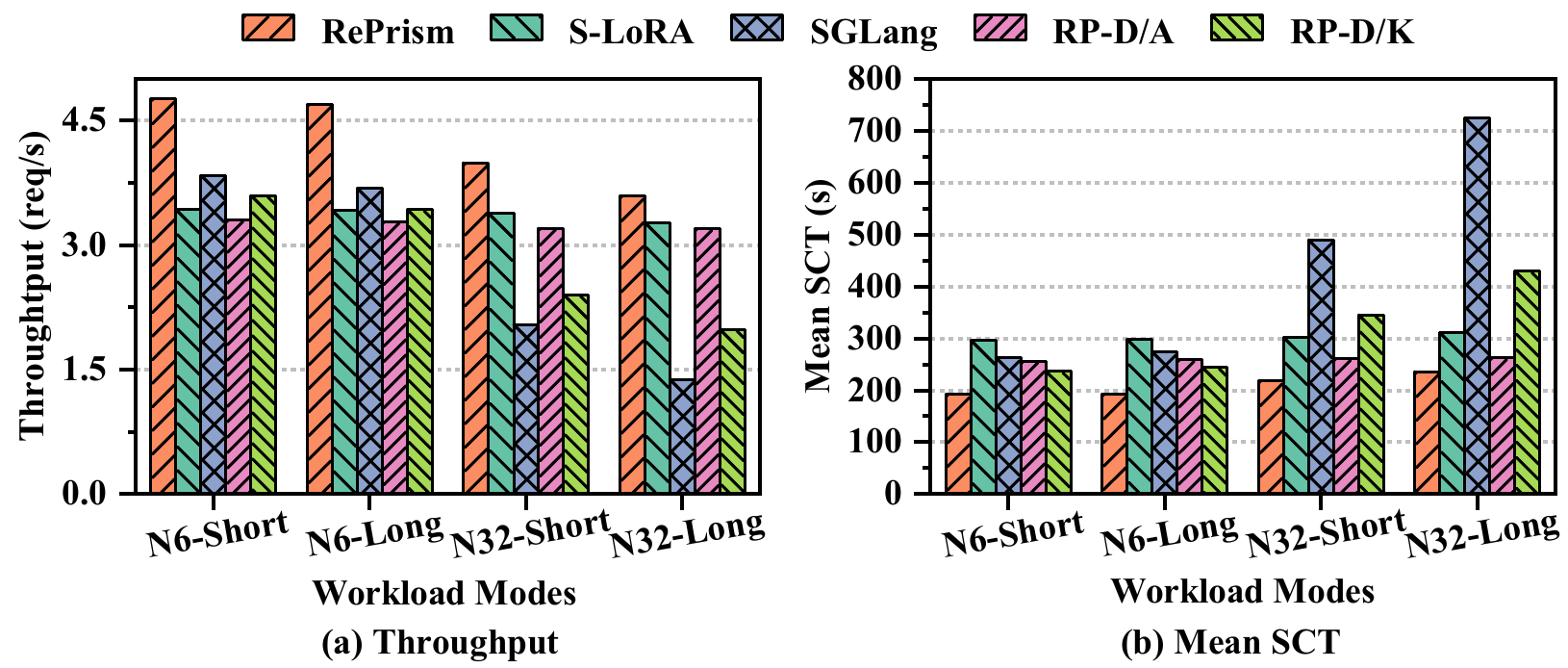}\par
\caption{Effect of reuse distance and the number of distinct \((P, R)\) pairs on (a) request throughput and (b) mean SCT under high concurrency. \(N\) counts distinct pairs. Short and Long denote reuse distances in the sequence of session root requests.}\label{fig:9}
\end{figure}

% word:103
\subsection{Performance across Session Arrival Rates}\label{sec:5.2}

% word:106
Figure~\ref{fig:8} compares the end-to-end performance of RePrism and the baselines on the dynamic workload. We omit P50 TTFT for Helium because it does not provide a timestamp for each request’s first token. RP-D/A and RP-D/K retain RePrism’s scoring mechanism but restrict the available representations to Direct/Adapter and Direct/KV, respectively. Matching the available representations isolates differences in execution order and state retention relative to the corresponding baselines.

% word:107
As the session arrival rate increases, SGLang and Helium approach capacity earlier than others. Their throughput grows more slowly while P95 SCT rises sharply. S-LoRA and Chameleon sustain higher arrival rates, but their throughput eventually levels off at about 10 to 11 requests/s while tail latency continues to rise. RePrism sustains about 14 requests/s under high load. At 1.8 sessions/s, it achieves 1.37\ensuremath{\times} the throughput of S-LoRA, the next-best system, and reduces P95 SCT by a factor of 4.10. Both the range of reuse options and the scheduling and cache management policies contribute to RePrism’s advantages. At high workloads, RP-D/A and RP-D/K also have lower P95 SCT than S-LoRA and SGLang, respectively These comparisons show that the policies retain some benefit even when the available representations are restricted. The advantage is more pronounced in P50 TTFT. At 1.7 sessions/s, RePrism reduces P50 TTFT by factors of 9.40 and 18.01 relative to S-LoRA and SGLang, respectively. RP-D/A also reduces P50 TTFT by a factor of 3.2 relative to S-LoRA.

% word:108
RePrism’s main gains are in throughput and SCT at medium to high loads. At higher loads, even with restricted representation choices, RP-D/A and RP-D/K achieve lower P95 SCT and substantially lower P50 TTFT than their respective baselines, S-LoRA and SGLang.

% word:119
\begin{figure*}[!t]
\centering
\includegraphics[width=.857\linewidth]{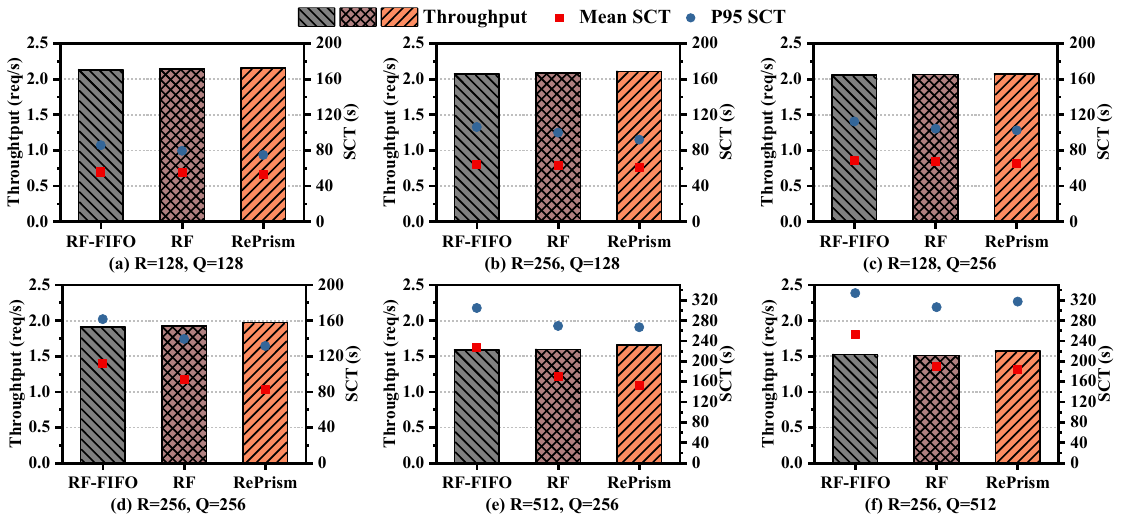}
\caption{Joint versus request-first scheduling across residual context lengths \(R\) and query lengths \(Q\). The \((R, Q)\) pairs, in tokens, are (a) (128, 128), (b) (256, 128), (c) (128, 256), (d) (256, 256), (e) (512, 256), and (f) (256, 512). Bars show throughput on the left axis. Squares and circles show mean and P95 SCT, respectively, on the right axis.}\label{fig:10}
\end{figure*}

% word:109
\subsection{Effects of Reuse Patterns under High Concurrency}\label{sec:5.3}

% word:110
We use a controlled workload to evaluate system performance under high concurrency. Figure~\ref{fig:9} compares four scenarios that vary in the number of distinct reusable \((P, R)\) pairs and their reuse distance in sessions. We denote the number of distinct pairs by \(N\) and set it to 6 or 32. Reuse distance is the number of intervening sessions between successive uses of the same pair. In Short, the same pair appears in consecutive sessions or a few sessions apart. In Long, sessions cycle through all \(N\) pairs. These settings apply only to the sequence of session root requests. During execution, the scheduler determines the order in which requests run and thus the actual reuse pattern.

% word:113
As Figure~\ref{fig:9} shows, RePrism achieves higher throughput and lower mean SCT than both S-LoRA and SGLang in all four scenarios. At \(N = 6\), speedups across the two metrics and both baselines range from about 1.24\ensuremath{\times} to 1.54\ensuremath{\times}. At \(N = 32\), SGLang and RP-D/K face greater resource pressure, and their performance degrades further at the longer reuse distance. Even so, RP-D/K reduces mean SCT by a factor of 1.69 relative to SGLang. The gains of RePrism and RP-D/A over S-LoRA narrow at \(N = 32\).

% word:114
RePrism’s advantage over the baselines is less pronounced on the high concurrency workload than on the dynamic workload. This difference is related to the configuration of request lengths used in the workload. In general, the larger the reusable fraction of each request, the more opportunity RePrism has to improve performance through scheduling.

% word:115
\subsection{Ablations and Policy Comparisons}\label{sec:5.4}

% word:116
\subsubsection{Effect of Joint Scoring}\label{sec:5.4.1}

% word:117
To evaluate the need for joint scoring, we compare RePrism with two policies that order requests before scoring each request’s eligible representations. The policies differ in how they order requests. RF uses the same request score as RePrism, while RF-FIFO serves requests in arrival order. Both rank representations in the same way as RePrism. The workload contains 32 distinct prompts \(P\) and runs in two consecutive phases. In each phase, half of the prompts are hot and appear repeatedly, while the other half are cold and appear only once each. The two groups switch roles in the second phase. We set the Adapter/Hybrid qualification coverage \(q\) to 50\%. Half of the prompts are eligible for Adapter and Hybrid, and the rest can use only Direct and KV.

% word:125
\begin{figure}[!t]
\centering
\includegraphics[width=.812\linewidth]{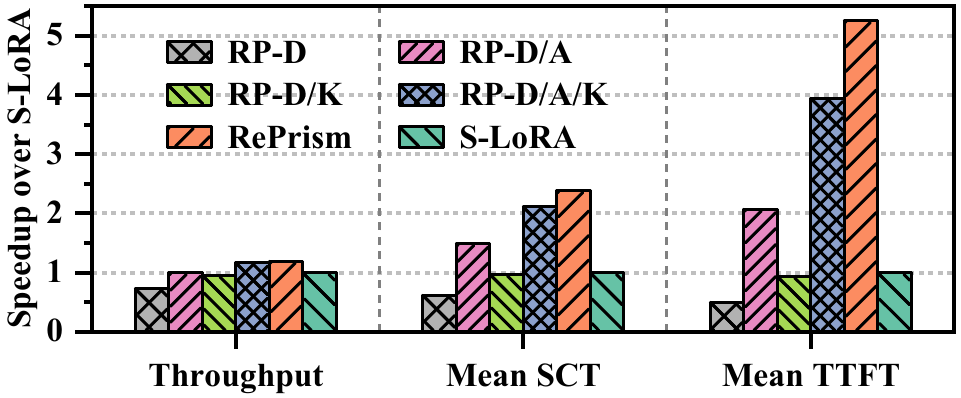}
\caption{Effect of the available prompt representations on throughput, mean SCT, and mean TTFT. All values are speedups over S-LoRA. Values above 1 indicate higher throughput or lower latency.}\label{fig:11}
\end{figure}

% word:120
Figure~\ref{fig:10} shows throughput and SCT for the three policies at different lengths of \(R\) and \(Q\). Across the tested lengths, RePrism’s throughput gains over RF-FIFO and RF are at most 5\%. The differences in mean and P95 SCT become more pronounced as \(R\) and \(Q\) grow longer. For example, at \(R = Q = 256\) tokens in Figure~\ref{fig:10}(d), RePrism reduces mean SCT by factors of 1.35 and 1.14 relative to RF-FIFO and RF, respectively. The corresponding reductions in P95 SCT are by factors of 1.23 and 1.06.

% word:121
At moderate lengths of \(R\) and \(Q\), ordering requests by score offers a substantial gain over FIFO, and RePrism’s joint scoring extends that gain. Scheduling offers limited gains when \(R\) and \(Q\) are short, while hardware limits lead to long request queues when \(R\) and \(Q\) are long. We therefore use \(R = Q = 256\) tokens in the subsequent experiments.

% word:122
\subsubsection{Available Prompt Representations}\label{sec:5.4.2}

% word:123
To evaluate how the choice of prompt representations affects RePrism’s performance, we construct four baselines with different sets of eligible representations. These baselines, RP-D, RP-D/A, RP-D/K, and RP-D/A/K, use the same scoring and resource management policies as RePrism.

% word:126
As Figure~\ref{fig:11} shows, RP-D uses Direct for every request and performs worst, with 27\% lower throughput than S-LoRA. Adding either Adapter (RP-D/A) or KV (RP-D/K) brings throughput close to that of S-LoRA. With more representations available, RP-D/A/K and RePrism both exceed S-LoRA’s throughput and reduce mean SCT by more than a factor of two. RePrism also reduces mean TTFT by a factor of 5.26 relative to S-LoRA. These results show that supporting multiple representations improves performance. In particular, avoiding repeated prefill of reusable prompt content can substantially reduce both SCT and TTFT.

% word:127
\subsubsection{Adaptive versus Fixed Representation Selection}\label{sec:5.4.3}

% word:128
In this section, we compare RePrism’s policy of choosing a representation at scheduling time with four policies that fix the representation in advance. S-D, S-A, S-K, and S-H always use Direct, Adapter, KV, and Hybrid, respectively, and otherwise follow RePrism’s scheduling policy. To highlight the effect of this choice, we construct two workloads, Cold and Hot, in which every prompt \(P\) can be represented by an adapter. Cold offers no opportunity for residual KV reuse, while Hot reuses 32 distinct residual KV caches.

% word:130
\begin{table}[!t]
\setlength{\belowcaptionskip}{4pt}
\caption{Mean SCT (s) with adaptive and fixed prompt representations under two reuse conditions.}\label{tab:adaptive-representations}
\centering
\setlength{\tabcolsep}{2.9pt}
\begin{tabular*}{\columnwidth}{@{}l@{\extracolsep{\fill}}rrrrr@{}}
\toprule
Reuse & S-D & S-K & S-A & S-H & RePrism \\
\midrule
Cold & 321.38 & 241.52 & \textbf{54.96} & 60.10 & 57.84 \\
Hot & 330.66 & 184.02 & 54.95 & \textbf{44.17} & 45.80 \\
\bottomrule
\end{tabular*}
\end{table}

% word:131
Table~\ref{tab:adaptive-representations} shows that the relative performance of S-A and S-H reverses between Cold and Hot, while S-D and S-K perform substantially worse in both. In Cold, S-A achieves a lower mean SCT than S-H (54.96 s versus 60.10 s). In Hot, S-H achieves a lower mean SCT than S-A (44.17 s versus 54.95 s). The best representation thus depends on the reuse conditions, and a single fixed choice cannot adapt to both. RePrism adjusts its choice at runtime, achieving a mean SCT of 57.84 s in Cold and 45.80 s in Hot. These values are only 5.2\% and 3.7\% higher, respectively, than those of the best static policy chosen in advance for each workload. Relative to that same baseline, always using Adapter incurs up to 24.4\% higher mean SCT in Hot, while always using Hybrid incurs up to 9.3\% higher mean SCT in Cold.

% word:132
\subsubsection{Cache Eviction Policies}\label{sec:5.4.4}

% word:133
We evaluate how the eviction policy affects RePrism’s performance using two variants, RP-LRU and RP-SE. Both differ from RePrism only in how they evict cache resources.

% word:135
\begin{figure}[!t]
\centering
\includegraphics[width=.95\linewidth]{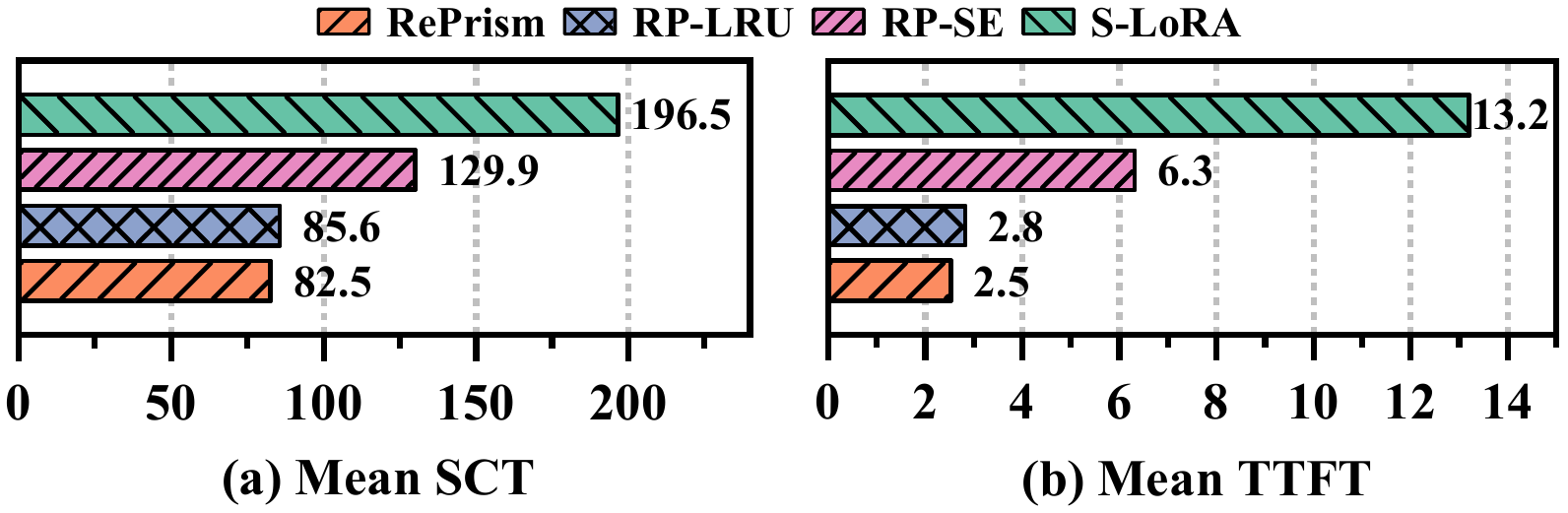}
\caption{Effect of cache resource eviction policy on (a) mean SCT and (b) mean TTFT.}\label{fig:12}
\end{figure}

% word:136
Figure~\ref{fig:12} compares mean SCT and mean TTFT across RePrism, RP-LRU, RP-SE, and S-LoRA. RP-SE discards cache resources once they are no longer in use. Its mean TTFT is lower than S-LoRA’s by a factor of 2.10 but remains much higher than that of RP-LRU or RePrism. When eviction is necessary, RP-LRU evicts the least recently used cache resources first. Under low GPU memory pressure, it can therefore retain resources needed by later requests. RP-LRU reduces mean TTFT by a factor of 4.68 relative to S-LoRA. RePrism uses workflow information to rank cache resources for eviction, so recency alone does not determine the eviction order. This reduces mean TTFT by a further factor of 1.13 relative to RP-LRU.

% word:137
\subsection{Sensitivity Analysis}\label{sec:5.5}

% word:138
\subsubsection{Choice of Base Model}\label{sec:5.5.1}

% word:140
\begin{figure}[!t]
\centering
\includegraphics[width=.858\linewidth]{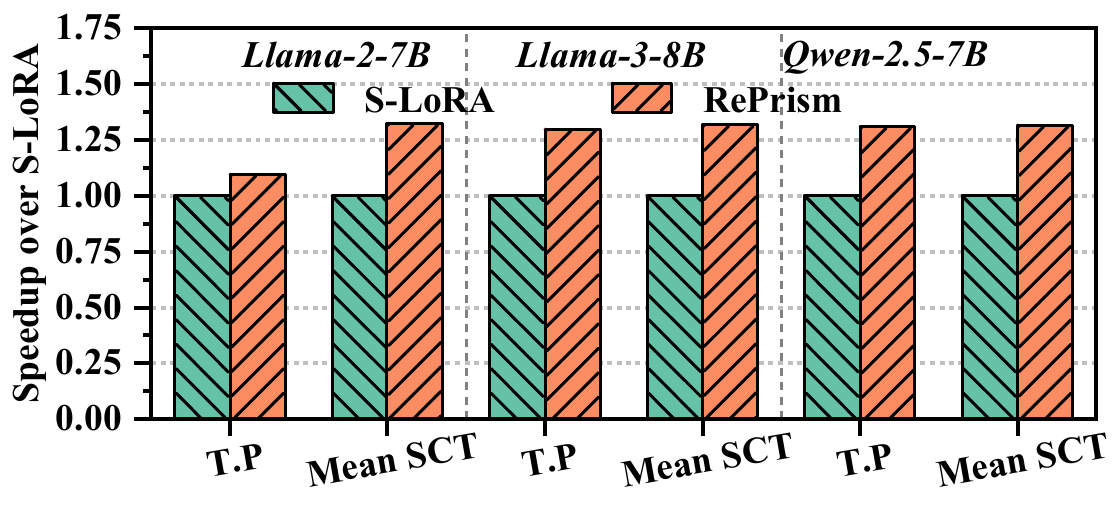}
\caption{RePrism’s throughput and mean SCT speedups over S-LoRA on the N32-Long workload. Each comparison uses S-LoRA on the same base model as the 1\ensuremath{\times} baseline.}\label{fig:13}
\end{figure}

% word:141
To evaluate how RePrism’s gains over S-LoRA vary with the base model, we test Llama-3-8B-Instruct~\cite{ref41} and Qwen-2.5-7B-Instruct~\cite{ref42} in addition to Llama-2-7B-Chat~\cite{ref43}. Because static GPU memory use varies across models, we calibrate the dynamic memory budgets to a common limit on peak GPU memory use. The resulting budgets are 8000 MiB for Llama-2, 5875 MiB for Llama-3, and 5687.5 MiB for Qwen-2.5. For each model, RePrism and S-LoRA use the same N32-Long workload and resource budget.

% word:145
\begin{figure}[!t]
\centering
\includegraphics[width=.80\linewidth]{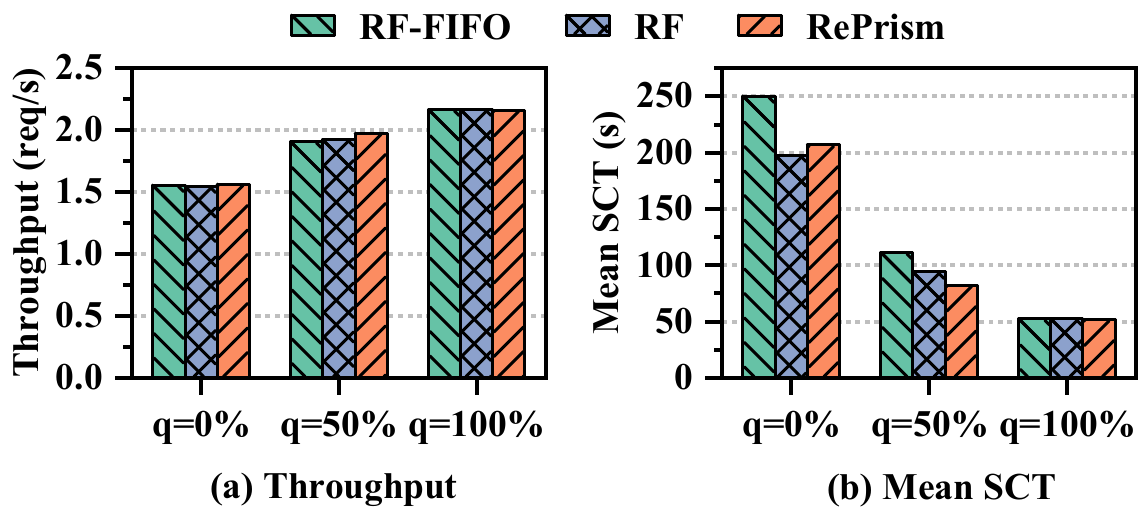}
\caption{Effect of adapter qualification coverage on (a) request throughput and (b) mean SCT. The coverage \(q\) is the fraction of distinct prompts eligible for Adapter and Hybrid.}\label{fig:14}
\end{figure}

% word:142
Figure~\ref{fig:13} compares throughput and mean SCT, with S-LoRA normalized to 1.0 for each model. RePrism achieves 1.10\ensuremath{\times}, 1.30\ensuremath{\times}, and 1.31\ensuremath{\times} the throughput of S-LoRA on Llama-2, Llama-3, and Qwen-2.5, respectively, and reduces mean SCT by a factor of 1.32 on all three. The larger throughput gains on Llama-3 and Qwen-2.5 are partly due to architectural differences. These models give the scheduler more capacity for reusable resources than Llama-2, allowing RePrism to cache more of those resources than S-LoRA. RePrism outperforms S-LoRA on all three models, suggesting that its gains extend beyond a single model architecture.

% word:143
\subsubsection{Adapter Qualification Coverage}\label{sec:5.5.2}

% word:146
To evaluate how sensitive RePrism’s performance is to the fraction \(q\) of distinct prompts \(P\) eligible for adapter internalization, we set \(q\) to 0\%, 50\%, and 100\% in this section.

% word:147
As Figure~\ref{fig:14} shows, increasing \(q\) from 0\% to 100\% raises RePrism’s throughput from 1.56 to 2.16 requests/s and reduces its mean SCT from 206.85 to 51.87 s. Joint scheduling offers its largest gain over request-first scheduling at \(q\) = 50\%, where RePrism achieves 2.7\% higher throughput and 12.3\% lower mean SCT than RF. At \(q\) = 100\%, RePrism reduces mean SCT by at most 1\% relative to RF. At \(q\) = 0\%, its mean SCT is slightly higher than RF’s. Thus, broader eligibility improves absolute performance on this workload, while joint request and representation selection offers a greater relative benefit when eligible and ineligible requests coexist.

% word:148
\subsubsection{Workflow Visibility}\label{sec:5.5.3}

% word:150
\begin{figure}[!t]
\centering
\includegraphics[width=.80\linewidth]{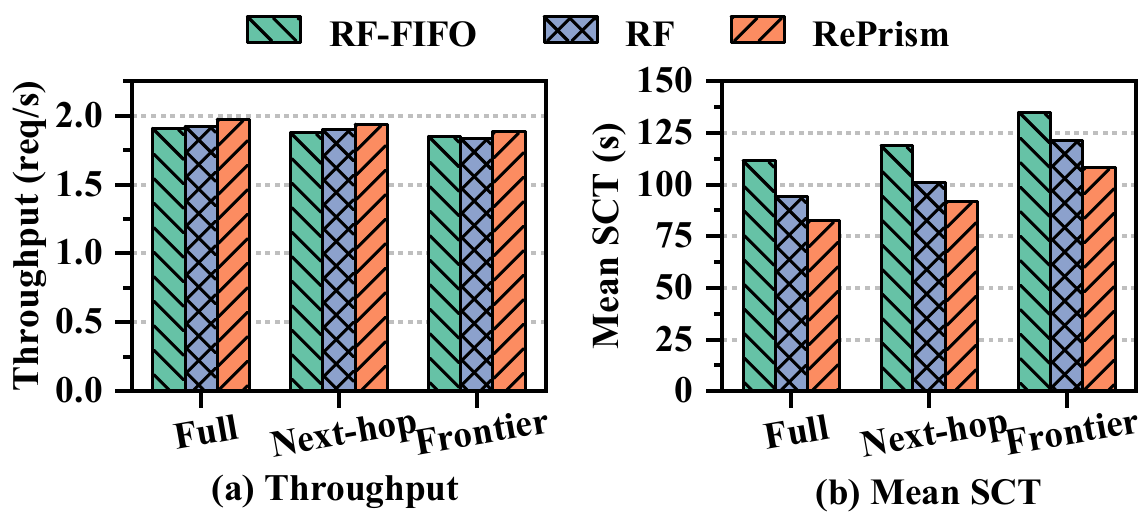}
\caption{Effect of workflow visibility on (a) request throughput and (b) mean SCT. Full exposes the entire submitted workflow, Next-hop exposes ready nodes and their immediate successors, and Frontier exposes only ready nodes.}\label{fig:15}
\end{figure}

% word:151
To evaluate how RePrism’s performance depends on which workflow nodes are visible to the scheduler, we test three visibility levels in this section. \textit{Full} exposes the complete workflow of each session the client has already submitted. \textit{Next-hop} exposes the currently ready nodes and their immediate successors. \textit{Frontier} exposes only the currently ready nodes. All three configurations use the complete workflow to enforce dependencies and release requests for execution. Only the information available to the scheduler varies.

% word:152
As Figure~\ref{fig:15} shows, reducing visibility from Full to Frontier lowers RePrism’s throughput from 1.97 to 1.89 requests/s and increases its mean SCT from 82.48 to 108.09 s. RePrism outperforms both RF and RF-FIFO at every visibility level. Relative to RF, it achieves 2.1\% to 2.7\% higher throughput and 9.1\% to 12.3\% lower mean SCT. Thus, reduced workflow visibility degrades RePrism’s performance, but it retains an advantage over both request-first policies.

% word:153
\section{Related Work}\label{sec:6}

% word:154
\textbf{Serving agent workflows.} Existing systems~\cite{ref06,ref07,ref08,ref09,ref44} use dependencies across calls to optimize agentic workflow execution. KVFlow~\cite{ref10} uses dependencies in its Agent Step Graph to predict upcoming calls and guide the eviction and prefetching of prefix KV states. Helium~\cite{ref21} models a workflow as a query plan to optimize reuse of cached data and computation across calls. SAGA~\cite{ref11} studies workflow-level scheduling. Continuum~\cite{ref45} uses a time-to-live policy to control KV retention during tool calls, and ScaleSim~\cite{ref39} uses call distance to manage agent state. RePrism further considers the eligible representations of each request, using workflow information to jointly select which ready requests to execute and which representation to use for each.

% word:155
\textbf{Prompt internalization.} Prior work~\cite{ref19,ref31,ref46,ref47} explores how to encode prompts or context in model parameters so that they need not be supplied explicitly at inference time. FIP~\cite{ref19} parameterizes fixed inputs, while Context Distillation~\cite{ref31} and GenPI~\cite{ref46} explore how to internalize the behavior induced by a prompt. Generative Adapter~\cite{ref47}, Doc-to-LoRA~\cite{ref20}, and Text-to-LoRA~\cite{ref48} generate lightweight adapters from context or task descriptions. Prompt compression methods~\cite{ref49,ref50,ref51,ref52} reduce prompt overhead by shortening the input or compressing its representation. RePrism treats previously constructed adapters that pass behavioral qualification as candidate representations. It studies online representation selection and resource management when these adapters coexist with explicit inputs and KV representations.

% word:156
\textbf{Adapter and KV serving.} Multi-LoRA serving systems~\cite{ref22,ref33,ref34,ref38,ref53,ref54} improve execution efficiency through heterogeneous batching, specialized operators, and adapter loading and caching. ELORA~\cite{ref23} exploits dependencies between LoRA and KV usage to manage both types of state in a unified cache pool. Work on KV serving~\cite{ref12,ref13,ref14,ref15,ref16,ref17,ref55,ref56,ref57} focuses on GPU memory allocation, prefix reuse, and caching across memory tiers. vLLM~\cite{ref16} uses PagedAttention to manage the KV cache, and SGLang~\cite{ref17} uses RadixAttention to reuse prefixes across requests. Recent work~\cite{ref58,ref59,ref60,ref61} also explores KV sharing across adapters or fine-tuned models. RePrism coordinates request scheduling and GPU memory allocation by accounting for how representation choice changes each request’s execution cost and the shared state it needs.

% word:158
\section{Conclusion}\label{sec:7}

% word:159
We present RePrism, a system that jointly schedules requests and selects their representations for agentic workflows. RePrism uses visible workflow information to rank candidate pairs of requests and representations for execution. Its Resource Manager coordinates the admission and eviction of reusable resources. On the dynamic workload at 1.8 sessions/s, RePrism achieves 1.37\ensuremath{\times} the request throughput of S-LoRA and reduces P95 SCT by a factor of 4.10. On the controlled workload, it reduces mean TTFT by a factor of 5.26 relative to S-LoRA. Joint scoring yields 2.7\% higher request throughput and 12.3\% lower mean SCT than ranking requests by execution score before choosing their representations.

\label{end-main-text}
\bibliographystyle{plain}
\AddToHookNext{cmd/bibitem/before}{\addtolength{\itemsep}{-1pt}}
\bibliography{references}
% 按正文原有方式平衡参考文献末页的双栏, 随后另起一页.
\flushcolsend
\onecolumn
\pdfbookmark[0]{Supplementary Material}{supplementary-material}
{\centering\Large\bfseries Supplementary Material\par}
\vspace{\baselineskip}

\newcommand{\samplecode}[1]{{\fontfamily{lmtt}\selectfont #1}}
\newcommand{\promptrule}[1]{\noindent #1\par\smallskip}
\setlength{\multicolsep}{4pt plus 1pt minus 1pt}

\begin{multicols}{2}\raggedcolumns

This supplement provides further details on the methods used in the paper. Section~A describes the scoring and resource management rules used in Section~4 and reports the CPU cost of batch selection. Section~B expands on the prompt internalization procedure in Section~3.

\appendix
\setcounter{equation}{0}
\setcounter{figure}{0}
\renewcommand{\theHequation}{supp.\arabic{equation}}
\renewcommand{\theHfigure}{supp.\arabic{figure}}
\renewcommand{\theequation}{A\arabic{equation}}
\renewcommand{\thefigure}{S\arabic{figure}}

\section{Scoring and Resource Management}

\subsection{Session Progress Value and Normalization}\label{supp:progress}

Section~4.3 computes the node execution value from two terms: the session progress value \(P_{t}(s,v)\) and the waiting term \(W_{t}(s)\), which reflects the time session \(s\) has spent waiting to be scheduled. \(P_{t}(s,v)\) gives higher priority to sessions nearing completion and, within a session, to nodes on longer remaining paths.

Let \(L_{v}\) be the maximum estimated time from node \(v\) to any terminal node in the visible workflow, including the time for \(v\) itself and all subsequent nodes on that path. Let \(L_{s}\) be the longest remaining path duration in the visible workflow of session \(s\). To compute \(L_{v}\), we first estimate each node's remaining execution time. For a node that has yet to run, we use its estimated execution time under Direct. For a running node, we scale that estimate by the fraction of its output budget remaining. For a node that has yet to reach the server, we also add its known tool wait time.

We then work backward from the terminal nodes. A terminal node's path duration equals its own estimated remaining time. For each other node, we add its own estimated remaining time to the largest path duration among its immediate successors. The maximum over all nodes gives \(L_{s}\). We define the session progress value as

\begin{equation}
  P_{t}(s,v)=\frac{H}{H+L_{s}}\cdot \frac{L_{v}}{L_{s}}
  \label{eq:progress}
\end{equation}

We set \(H\) to 60 seconds to establish the time scale. The first ratio increases as \(L_{s}\) decreases, giving higher progress values to sessions with shorter remaining paths. Within a session, the second ratio, \(\frac{L_{v}}{L_{s}}\), gives higher progress values to nodes with longer remaining paths.

We introduce the waiting term \(W_{t}(s)\) to prevent sessions from waiting too long. Waiting time \(w_{s}\) accumulates only while the session has at least one ready request and no running requests. We compute \(W_{t}(s)=\min \left(1,\frac{w_{s}}{H}\right)\), so time spent executing requests does not increase this term. We then normalize the node execution value and the representation value separately to \([0.5, 1]\) and multiply them. This prevents either value from dominating the ranking because its raw score is unusually large or small.

\subsection{Cost Estimates for Representation Scoring}\label{supp:scoring}

Section~4.3 breaks the representation value into the immediate execution benefit \(V_{t}^{\mathrm{now}}(v,m)\), the future reuse benefit \(G_{t}^{\mathrm{reuse}}(v,m)\), and the resource opportunity cost \(O_{t}^{\mathrm{resource}}(v,m)\). To compute \(V_{t}^{\mathrm{now}}(v,m)\), we first estimate the remaining execution time of the same request under Direct and under representation \(m\), denoted by \(T_{D}\) and \(T_{m}\), respectively. Their difference gives the execution time saved by using \(m\). Let \(\Delta T_{\mathrm{load}}\) denote the additional time needed to prepare \(m\). The immediate execution benefit is

\begin{equation}
  V_{t}^{\mathrm{now}}(v,m)=\frac{T_{D}-T_{m}-\Delta T_{\mathrm{load}}}{T_{D}}
  \label{eq:execution}
\end{equation}

Both \(T_{D}\) and \(T_{m}\) include prefill and decoding. To compute \(\Delta T_{\mathrm{load}}\), RePrism compares the resource preparation plans before and after adding the candidate and counts only the additional time needed to load or build resources for \(m\). These time estimates come from offline profiling. When the lengths match a measured configuration, RePrism can use the corresponding measurements directly. Otherwise, it estimates the times from the actual input length, the output budget, and the adapter rank.

The future reuse benefit \(G_{t}^{\mathrm{reuse}}(v,m)\) captures the value of later requests reusing the current representation. The scheduler considers only pending requests visible in the workflow. It excludes the current request and those already selected for the batch in this round to avoid counting the same benefit twice. Let \(\Delta T_{u}\) be the execution time that later request \(u\) is expected to save through reuse. The future reuse benefit is

\begin{equation}
  G_{t}^{\mathrm{reuse}}(v,m)=\frac{1}{T_{D}}\sum_{u} \frac{p_{u}\,\Delta T_{u}}{1+d_{u}/T_{0}}
  \label{eq:reuse}
\end{equation}

Here, \(p_{u}\) is the estimated probability that request \(u\) will execute. RePrism sets \(p_{u}\) to 1 for requests already submitted and uses the known branch probability for the remaining requests. The term \(d_{u}\) is the estimated time until the execution dependencies of \(u\) are satisfied. We set \(T_{0}\) to 1 second to establish the time scale for discounting. Requests that are more likely to execute and whose dependencies will be satisfied sooner therefore contribute more to the future reuse benefit. When the same later request can benefit from multiple shared resources, RePrism distributes the total time savings across those resources.

Adding a candidate can also put pressure on GPU memory and evict resident cache resources. The resource opportunity cost \(O_{t}^{\mathrm{resource}}(v,m)\) measures the additional cost of restoring those resources. RePrism estimates which resources would be evicted before and after adding the candidate and compares the costs of rebuilding them. It divides the difference by \(T_{D}\) and clips the result to \([-1, 1]\) to obtain the cost term used in scoring. The rebuild cost estimates also come from offline profiling.

\subsection{Batching and Peak Dynamic KV Usage}\label{supp:peak}

Section~4.4 checks the GPU memory required when the proposed batch is merged with the running batch. This requirement depends on how prefill is performed and how dynamic KV usage changes during decoding. Each generated token increases KV usage, while shorter requests release their memory as they finish. Current usage alone therefore does not capture the batch peak, and individual requests can reach their peaks at different times. We sort the requests in the batch by remaining output budget in descending order. The batch's peak KV usage during decoding is

\begin{equation}
  T_{\mathrm{decode}}=\max_{k}\left(\sum_{i=1}^{k} u_{i}+k\,o_{k}\right)
  \label{eq:decode-peak}
\end{equation}

Here, \(u_{i}\) is the number of dynamic tokens currently held by request \(i\), and \(o_{i}\) is its remaining output budget. We evaluate the expression for each \(k\) from 1 to the number of requests and take the maximum as \(T_{\mathrm{decode}}\). RePrism uses Equation~\eqref{eq:decode-peak} to estimate the dynamic KV requirement during decoding and also accounts for memory use during prefill when checking GPU memory capacity.

\subsection{Scheduling Overhead}

We first populate the ready queue, then time one call to RePrism's batch selection routine on the CPU. Each request belongs to a separate session with one node and can use all four representations. Figure~\ref{fig:scheduling-overhead} shows the resulting latency for 1 to 128 ready requests, or 4 to 512 candidates.

Median latency rises from 1.01\,ms at four candidates to 17.19\,ms at 128 and 62.45\,ms at 512. At 512 candidates, P95 and P99 reach 64.21 and 65.64\,ms, respectively.

\par\smallskip\noindent\begin{minipage}{\linewidth}
  \makeatletter
  \def\@captype{figure}%
  \makeatother
  \setlength{\abovecaptionskip}{5pt}
  \centering
  \includegraphics[width=.78\linewidth]{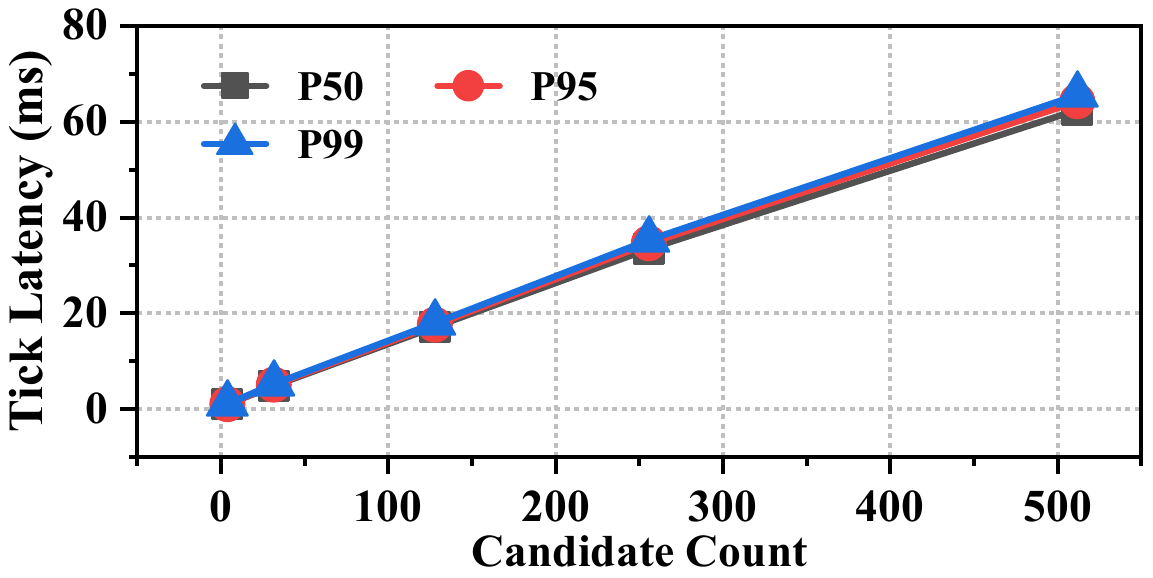}
  \LinkTargetOff % 图题会创建图 S1 的跳转目标.
  \caption{CPU latency per scheduling round (tick). Each percentile is computed over 1000 ticks per queue size and averaged across five runs.}
  \label{fig:scheduling-overhead}
\end{minipage}\par\smallskip

\columnbreak
\section{Training and Behavioral Checks for Prompt Internalization}\label{supp:training}

We use the Media\_1 state tracking task from the Schema-Guided Dialogue (SGD) dataset to illustrate the adapter training procedure described in Section~3.1. In this task, \(P\) specifies the state update rules, \(R\) provides the previous action and known information, and \(Q\) expresses the current request. The target answers are derived from the SGD annotations. We train on Llama-2-7B-Chat with the base model weights fixed. We update only the LoRA parameters in the attention layers, using the model's prediction error on the target answers. The training example in Appendix~\ref{app:example} shows how the fixed prompt and dynamic input determine the target output. Direct receives both \(P\) and the dynamic input. With the adapter loaded, the model receives only \(R\) and \(Q\).

In this example, the current request asks to enable subtitles, so \texttt{SUBTITLES} is set to \texttt{True}. The intent remains \texttt{PlayMovie}, so the title is carried over from \(R\), while the director and genre fields are set to \texttt{NONE} because they do not apply to this action. The target output thus reflects both the new request and the relevant prior state. The training and development sets contain 48 and 12 examples, respectively, with a separate set of 24 examples for behavioral qualification. During evaluation, we select the most probable token at each step and generate up to 64 new tokens to check whether the adapter follows the rules on these held-out examples. We first check that all five fields are present, then compare the predicted intent and the required field values against the target. The behavioral score is the fraction of examples that pass these checks. Field order may vary, and outputs may differ token by token from Direct as long as they satisfy the predefined behavioral rules. If an adapter falls short of the required behavioral score and its current rank is below the preset threshold, we increase the rank, retrain the adapter, and repeat the checks. We select the lowest rank that passes.

\end{multicols}
\medskip\noindent\begin{minipage}{\textwidth}
\section{Media\_1 State Tracking Example}\label{app:example}
\hrule height .5pt\vspace{5pt}
\noindent\begin{minipage}[t]{.173\linewidth}
\textbf{Fixed prompt \(P\)}\\(excerpt)
\end{minipage}\hfill\begin{minipage}[t]{.80\linewidth}
\raggedright
\promptrule{Track \samplecode{Media\_1} state from \samplecode{CURRENT\_UTTERANCE}, \samplecode{PREVIOUS\_INTENT}, and \samplecode{PREVIOUS\_SLOTS}.\\Return only the updated state.}
\promptrule{Use \samplecode{CURRENT}\textquotesingle{}s most specific cue: subtitle(s) \samplecode{-\textgreater{}} \samplecode{PlayMovie}; \samplecode{...}}
\promptrule{For the same intent, carry relevant \samplecode{PREVIOUS} fields not replaced by \samplecode{CURRENT},\\character for character.}
\promptrule{\samplecode{SUBTITLES} requires the literal word subtitle(s).\\With/\allowbreak{}on/\allowbreak{}add/\allowbreak{}enable/\allowbreak{}display/\allowbreak{}leave on/\allowbreak{}need/\allowbreak{}want means \samplecode{True}; \samplecode{...}}
\promptrule{\samplecode{PlayMovie} forces \samplecode{DIRECTED\_BY=NONE} and \samplecode{GENRE=NONE}.}
Output exactly five non-empty \samplecode{FIELD=value} lines, each once. Order is irrelevant.
\end{minipage}\par
\vspace{5pt}\hrule height .25pt\vspace{5pt}
\noindent\begin{minipage}[t]{.173\linewidth}
\textbf{Dynamic input}\\\(R+Q\)
\end{minipage}\hfill\begin{minipage}[t]{.80\linewidth}
\raggedright
\samplecode{CURRENT\_UTTERANCE=Actually, please play with subtitles.}\\
\samplecode{PREVIOUS\_INTENT=PlayMovie}\\
\samplecode{PREVIOUS\_SLOTS=title=Body Double}
\end{minipage}\par
\vspace{5pt}\hrule height .25pt\vspace{5pt}
\noindent\begin{minipage}[t]{.173\linewidth}
\textbf{Expected output}
\end{minipage}\hfill\begin{minipage}[t]{.80\linewidth}
\raggedright
\samplecode{INTENT=PlayMovie}\\
\samplecode{DIRECTED\_BY=NONE}\\
\samplecode{GENRE=NONE}\\
\samplecode{SUBTITLES=True}\\
\samplecode{TITLE=Body Double}
\end{minipage}\par
\vspace{5pt}\hrule height .5pt
\end{minipage}

\end{document}